\documentclass[superscriptaddress,aps,pra,twocolumn,showpacs,nofootinbib,longbibliography,notitlepage]{revtex4-1}
\usepackage{etex}
\usepackage{amsmath,amssymb,amsthm}
\usepackage[colorlinks=true,citecolor=blue,urlcolor=blue]{hyperref}
\usepackage[pdftex]{graphicx}
\usepackage{times,txfonts}
\usepackage{braket}
\usepackage{color}
\usepackage{natbib}
\usepackage{amsmath,blkarray}
\usepackage{mathtools}
\usepackage{latexsym}
\usepackage{tabularx, booktabs}
\usepackage{graphics,epstopdf}
\usepackage{graphicx}
\usepackage{float}
\usepackage{graphicx}
\usepackage{amsfonts}
\usepackage{subcaption}
\usepackage{color,soul}

\newcommand{\be}{\begin{equation}}
\newcommand{\ee}{\end{equation}}
\newcommand{\ba}{\begin{eqnarray}}
\newcommand{\ea}{\end{eqnarray}}

\begin{document}
\title{Revealing hidden nonlocality and preparation contextuality  for an arbitrary input Bell inequality}
\author{ Asmita Kumari }
\email{asmita.physics@gmail.com}
\affiliation{ CSIR-National Physical Laboratory, Dr. K. S. Krishnan Road, New Delhi 110012, India. }
\affiliation{Department of Mathematical Sciences, Indian Institute of Science Education and Research Berhampur, Berhampur 760010, Odisha, India}
\author{ Saikat Patra }
\affiliation{Department of Mathematical Sciences, Indian Institute of Science Education and Research Berhampur, Berhampur 760010, Odisha, India}

\begin{abstract}
 In recent years, the activation of hidden nonlocality for a mixed entangled state, admitting a local model, has gained considerable interest. In this paper, we study the activation of hidden nonlocality and preparation contextuality for a class of mixed entangled states, using local filtering operations. For our demonstration, we consider the two-party (Alice and Bob) one-way communication game known as parity oblivious random access code (PORAC). The quantum success probability of such $n$-bit PORAC solely depends on a Bell functional involving $2^{n-1}$ and $n$ dichotomic measurement settings for Alice and Bob, respectively. Such a Bell functional has two classical bounds, the local and the preparation non-contextual. We show that using local filtering operations on local mixed entangled state, the nonlocality can be revealed for any non-zero value of the mixedness parameter of the entangled state if $n\geq6$. Further, we show that the preparation contextuality, which is a comparatively weaker quantum correlation than nonlocality, can be revealed for any non-zero value of the mixedness parameter of the mixed entangled state if $n\geq 4$.
 \end{abstract}

\maketitle

\section{Introduction}
Bell's theorem \cite{Bell1964} is one of the most fundamental results in quantum foundations that demonstrates that any theory satisfying local realism cannot mimic quantum theory. This feature is widely known as nonlocality and commonly demonstrated through the quantum violation of a suitable Bell inequality. Another celebrated non-classical feature, the incompatibility between the non-contextual model and quantum theory was demonstrated by the Kochen and Specker (KS) \cite{kochen67} theorem. Besides adding enormous importance in quantum foundations both nonlocality and contextuality have been proven to be resources in developing quantum technologies, such as in communication games \cite{complx1,mig}, randomness certification \cite{col06,pir10,nieto,col12,roch}, secure quantum key distribution \cite{bar05,acin06,acin07,pir09}, witnessing Hilbert space dimension \cite{wehner,gallego,ahrens,brunnerprl13,bowler,sik16prl,cong17,pan2020} and quantum computation \cite{vega,frem,brav}. 

The traditional KS contextuality has limited scope because it is defined using the assumptions of outcome determinism for sharp measurement and measurement non-contextuality. Spekkens in 2005 \cite{spe05} generalized the formulation of KS contextuality for unsharp measurement and also extended it to preparation and transformation non-contextuality. In this paper, this generalized notion of contextuality is considered, and we demonstrate the activation of preparation contextuality.

The entanglement \cite{Schrödinger1935,epr1935,horodecki09} between specially separated quantum systems is necessary to realize nonclassical features such as nonlocality and contextuality assisted by entanglement \cite{ghorai18,kumari2019,kumari2023}. However, entanglement is not sufficient to generate such correlations. There exist entangled states that admit local models. Therefore, the degree of entanglement plays a key role.  Popescu \cite{Popescu1995} has demonstrated a key approach to activate nonlocality by using local filtering operation even when the entangled state is demonstrating locality. 
Since then, a multitude of works have been reported to reveal hidden nonlocality for various types of entangled states admitting local models \cite{hao,ku,lli,jones,Popescu1995,GISIN1996,asmita24,lli,fang24}. In this regard, we note that in recent times the study of contextuality correlation has gained considerable attention \cite{tava}. It has been demonstrated \cite{ghorai18} that even if a quantum correlation cannot demonstrate nonlocality, it can still demonstrate contextuality, a somewhat weaker correlation than nonlocality.  

In this paper, by applying suitable local filtering operations, we reveal the hidden nonlocality and preparation contextuality for a mixed entangled state admitting a local model. 
In order to demonstrate this, we consider a two-party (Alice and Bob) communication game known as $n$-bit PORAC. The success probability of that game solely dependent on a Bell functional \cite{ghorai18} featuring $2^{n-1}$ and $n$ dichotomic measurements of Alice and Bob, respectively, where $n$ is arbitrary. For any value of $n$, such a Bell functional has a preparation non-contextual bound and a local bound, which is always higher except for $n=2$. Here, the assumption of preparation non-contextuality originates from the functional relation between Alice's  observables, leading to a constrained scenario and eventually leading to a lower value than the local bound. We show that the nonlocality can be activated for the whole range of mixedness parameter $q\in[0,1\}$ if $n \geq 6$. However, preparation contextuality can be revealed for any nonzero value of the mixing parameter if $n \geq 4$.

This paper is organized as follows. In Sec. II, we provide a brief description of the ontological model of an operational theory, the success probability of $n$-bit PORAC in local and preparation non-contextual models, and in quantum theory. In Sec. III, we examine the restrictions on the mixedness parameter of mixed entangled state for observing nonlocality and preparation contextuality using the optimal quantum violation of local and preparation non-contextual bound of a family of Bell inequalities. Sec. IV demonstrates the application of suitable local filtering operation on mixed entangled states for revealing hidden nonlocality and preparation contextuality using the quantum violation of respective classical bound of Bell inequalities for $n=2,3,4$ and $5$. We generalize the effect of the local filtering operation for arbitrary $n$ in Sec. V. Finally in Sec. VII, we summarize our findings.
\section{Preliminaries}
To begin with, we summarize the ontological model of an operational theory and notion of non-contextuality. Then we discuss the success probabilities of PORAC in local and preparation non-contextual models. In the last subsection of the preliminary, the advantages in success probabilities of PORAC in quantum theory are discussed.  
\subsection{Ontological model and the notion of non-contextuality}
 The framework of the ontological model of an operational theory was introduced in  \cite{har10,spe05} which can be described as follows \cite{har10,spe05}. For the case of quantum theory, whenever the density matrix $\rho$ is prepared by the preparation procedure $P$, it is assumed that in an ontological model a probability distribution $\mu_{P}(\lambda|\rho)$ of the ontic state $\lambda\in \Lambda$ is prepared, satisfying $\int _\Lambda \mu_{P}(\lambda|\rho)d\lambda=1$. Here, $\Lambda$ is the ontic state space. In quantum theory, the measurement is in general described by positive-operator-valued-measure $E_{k}$ with $\sum_{k} E_k=\mathbb{I}$. In an ontological model, when the measurement of $E_{k}$ is performed through a measurement procedure $M$,  the ontic state assigns the response function $\xi_{M}(k|\lambda, E_{k}) $, satisfying $\sum_{k}\xi_{M}(k|\lambda , E_{k})=1$. An ontological model must reproduce the Born rule, i.e., Hence, $\forall \rho $, $\forall E_{k}$ and $\forall k$, $\int _\Lambda \mu_{P}(\lambda|\rho) \xi_{M}(k|\lambda, E_{k}) d\lambda =Tr[\rho E_{k}]$.

 Inspired by Leibniz's principle, the concept of non-contextuality in an ontological model emerges from the equivalence class of experimental procedures. If two preparation procedures $P$ and $P^{\prime}$ prepare the same density matrix $\rho$, then no measurement can operationally distinguish the context by which $\rho$ is prepared. Therefore, they can be considered as equivalent preparations.  Such an equivalence leads to the notion of preparation non-contextuality assumption in an ontological model of quantum theory, i.e.,  
\begin{align}
\label{ass}
\forall M, \  k: \ \   p(k|P, M)=p(k|P^{\prime},M)	\Rightarrow \mu_{P}(\lambda|\rho)=\mu_{P^{\prime}}(\lambda|\rho)
\end{align}
 This simply means that the assumption of preparation non-contextuality dictates that the ontic state distributions are equivalent in two preparation contexts $ P $ and $ P^{\prime}$ \cite{spe05,akp19,pan2021}. 
 
\subsection{Success probabilities of PORAC in local and preparation non-contextual models}
A $n$-bit PORAC \cite{spek09} is a two-party one-way communication game that can be summarized as follows. Let us assume Alice as a sender chooses $n$-bit string $x$ randomly from $x\in\{0,1\}^{n}$ and sends it to Bob. Bob receives $ y \in \{1,2, ..., n\}$ as inputs uniformly at random and produces output $b$. The task of Bob is to recover the ${y}^{th}$ bit of Alice's input string with a probability. Both Alice and Bob try to optimize the success probability of one-way communication game with the condition of $b=x_y$. The average success probability of winning the game is defined as
	\begin{equation}
		\label{qprob}
		p = \dfrac{1}{2^n n}\sum\limits_{x,y}p(b=x_y|x,y).
	\end{equation}
In order to maximize the success probability, Alice is allowed to share an unbounded amount of information to Bob with a restriction known as parity-oblivious constraint. Parity-oblivious constraints on the information sent by Alice to Bob demonstrate that Alice is allowed to share information with Bob in such a way that information of parity should not be known to Bob.  

Spekkens \emph{et al.} \cite{spek09} first defines the parity-oblivious constraint for any $s$ belonging to the parity set $ \mathbb{P}_n= \{x|x \in \{0,1\}^n,\sum_{r} x_{r} \geq 2\} $ with $r\in \{1,2,...,n\}$ as no information of $s\cdot x = \oplus_{r} s_{r}x_{r}$ (s-parity) should be transmitted to Bob. Such a restriction allows for sending only one bit of information. The classical bound of success probability for $n$-bit RAC \cite{spek09} is given by 
\begin{align}
		\label{cb}
	p_{pnc}	 \leq \frac{1}{2}\left(1+\frac{1}{n}\right) 
	\end{align}
It has been argued \cite{spek09,akp19,kumari2019} that if the ontological model of the quantum theory is preparation non-contextual then the parity-oblivious constraint at the operation level is satisfied at the level of the ontic state. Hence the classical bound of success probability in Eq.(\ref{cb}), obtained using parity-oblivious constraint at the operational level is a preparation non-contextual bound at the ontological level. The form of non-classicality observed through the quantum violation of preparation non-contextual bound in Eq.(\ref{cb}) is preparation contextuality.

In quantum PORAC, Alice encodes her $n$-bit string into quantum states $ \rho_x $ and sends it to Bob. Here, it should be noted that we are considering an entanglement-assisted $n$-bit RAC communication game, in which Alice and Bob are allowed to share a suitable entangled state. Then Alice can steer the quantum states $ \rho_x $ available to Bob by performing $2^{n-1}$ measurements. After receiving the state from Alice, Bob performs dichotomic measurement for every $y \in \{1,2,...,n\}$ and announces output $b$. For this case, the quantum success probability transform \cite{ghorai18} as
\begin{align}
\label{qprobn0}
p_Q = \dfrac{1}{2} + \dfrac{1}{2^n n}\sum_{y=1}^{n} \sum\limits_{i=1}^{2^{n-1}} (-1)^{x^i_y}  \langle A_{n,i}\otimes B_{n,y}\rangle 
\end{align}
From Eq.(\ref{qprobn0}), it can be seen that the quantum success probability depends on the quantum value of the Bell expression 
\begin{align}
\label{nbell1}
	\mathcal{B}_{n} =  \sum_{y=1}^{n}\sum_{i=1}^{2^{n-1}} (-1)^{x^i_y}  A_{n,i}\otimes B_{n,y} 
\end{align}
 Note that for $n=2$ and $3$, $\mathcal{B}_{n}$ transform to CHSH \cite{clause1969} and Gisin's elegant Bell \cite{gisin} expressions respectively. 

Following \cite{kumari2019} the local bound of Bell expression in Eq.(\ref {nbell1}) is given by
\begin{align} 
\label{localbound}
	\mathcal{B}_{n,L} \leq n\binom{n-1}{\lfloor\frac{n-1}{2}\rfloor}  
\end{align}
However, the preparation non-contextuality bound is obtained by imposing the parity oblivious condition in quantum theory, which demands 
	\begin{align}
		\label{poc}
		\forall s: \frac{1}{2^{n-1}}\sum\limits_{x|x.s=0} \rho_{x}=\frac{1}{2^{n-1}}\sum\limits_{x|x.s=1} \rho_{x}
	\end{align}
This parity oblivious constraint puts a condition on the Alice observable as given below
\begin{equation}
\label{ntpnc}
\forall s: \sum_{i=1}^{2^{n-1}} (-1)^{s.x^i} A_{n,i} = 0
\end{equation}
Using this condition on Alice's observable, the local bound of the Bell expression reduces to the preparation non-contextual bound given by
\begin{align}
\label{npnc}
	{\mathcal{B}}_{n,pnc}\leq  2^{n-1}
\end{align}
By comparing local bound in Eq.(\ref{localbound}) and  preparation non-contextual bound in Eq.(\ref{npnc}) it can be seen that ${\mathcal{B}}_{n,L}  \geq	{\mathcal{B}}_{n,pnc}$. The equality holds only for $n=2$. Therefore, there may be a scenario in which nonlocality cannot be demonstrated, but there is still a chance to demonstrate preparation contextuality.

\subsection{Optimal quantum value of the Bell expression in Eq.(\ref{nbell1})}
Following \cite{ghorai18}, the optimal quantum value of Bell expression in Eq.(\ref{nbell1}) is given by
\begin{align}
\label{maxq}
\mathcal{B}^{opt}_{n,Q} = 2^{n-1}\sqrt{n}
\end{align}
which is achieved \cite{ghorai18} when Alice and Bob shares $m=\lfloor\frac{n}{2}\rfloor$ number of two-qubit maximally entangled state $|\phi_{2^{m}} \rangle =|\phi_2 \rangle^{\bigotimes m} $ where 
\begin{eqnarray}
|\phi_2 \rangle = \frac{1}{\sqrt{2}}(|0 \rangle_A|0 \rangle_B +|1 \rangle_A|1 \rangle_B)
\end{eqnarray}
Also, Bob's observables are mutually anti-commuting and Alice's observables satisfy the following relation  
\begin{equation}
\label{abr}
\sum_{i=1}^{2^{n-1}} (-1)^{s.x^i_{y}} A_{n,i} \otimes I = \frac{2^{n-1}}{\sqrt{n}} I \otimes B_{n,y}.  
\end{equation}

The set of observable of Bob is given in Table I. We note that the optimal quantum value of the Bell expression needs higher dimensional system to achieve.  For $n=2$ and $n=3$, the required  shared state is a two-qubit maximally entangled state $|\phi_2 \rangle $ and for  $n=4$ or $n=5$, Alice and Bob need to share a pair of two-qubit maximally entangled state is of the form 
\begin{eqnarray}
\label{s22}
|\phi_4 \rangle &=&|\phi_2 \rangle^{\bigotimes 2} \\ \nonumber &=&\frac{1}{2}\bigg[|00 \rangle_{A}|00 \rangle_{B} +|01 \rangle_{A}|01 \rangle_{B} +|10 \rangle_{A}|10 \rangle_{B}+|11 \rangle_{A}|11 \rangle_{B} \bigg] 
\end{eqnarray}
For the convenience of notation and for using local filtering operation, the state can be rewritten as
\begin{eqnarray}
\label{s4}
|\phi_4 \rangle = \frac{1}{2}\bigg[|0 \rangle_{A}|0 \rangle_{B} +|1 \rangle_{A}|1 \rangle_{B} +|2 \rangle_{A}|2 \rangle_{B}+|3 \rangle_{A}|3 \rangle_{B} \bigg] 
\end{eqnarray}
where we use notations 
\begin{align}
|00 \rangle \rightarrow |0 \rangle, |01 \rangle \rightarrow |1 \rangle, |10 \rangle \rightarrow |2 \rangle \  \  and \  \ |11 \rangle \rightarrow |3 \rangle
\end{align}
Similarly, for any arbitrary $n$,  $m=\lfloor\frac{n}{2}\rfloor$ with the notation  
\begin{eqnarray}
\nonumber 
&&|0 \rangle^{\bigotimes m} \rightarrow |0 \rangle, |0 \rangle^{\bigotimes m-1}|1 \rangle \rightarrow |1 \rangle, |0 \rangle^{\bigotimes m-2}|10 \rangle \rightarrow |2 \rangle, \\ \nonumber && |0 \rangle^{\bigotimes m-3}|100 \rangle \rightarrow |3 \rangle,......, |1 \rangle^{\bigotimes m} \rightarrow |2^m-1 \rangle
\end{eqnarray}
the state can be rewritten as
\begin{eqnarray}
\label{k}
|\phi_{2^{m}} \rangle &=&  \bigg(\frac{1}{ \sqrt{2} } \bigg)^m\sum_{k=0}^{2^{m}-1} |K \rangle_A|K \rangle_B
\end{eqnarray}
where $K=0,1,2,...n$.

\begin{table}
    \begin{tabular}{|c|c|}
    \hline
\multicolumn{2}{|c|}{Suitable choice of observable to get optimal quantum value}\\
    \hline
     Bit    &  $B_{n,y}$    ($y=1,2,3,.....,n$)   \\
     \hline
      2  &   $B_{2,1} = \sigma_x$, $B_{2,2} = \sigma_y$ \\
      \hline
      3   &  $B_{3,1} = \sigma_x$, $B_{3,2} = \sigma_y$, $B_{3,3} = \sigma_z$  \\
      \hline
      4   & $B_{4,1}=\sigma_x  \otimes\sigma_x $, $B_{4,2} =\sigma_x  \otimes \sigma_y$, $B_{4,3} = \sigma_x \otimes\sigma_z$,$B_{4,4} = \sigma_y \otimes I$ \\
      \hline
      \small{Odd $n$}  & $B_{n,y}=\sigma_x \otimes B_{n-2,y}$ ($y=1,.,n-2$),$B_{n,n-1} = \sigma_y \otimes I$,$B_{n,n} = \sigma_z \otimes I$   \\  
       \hline
      \small{Even $n$}   & $B_{n,y} = \sigma_x  \otimes  B_{n-1,y}$ ($y=1,2..,n-1$), $B_{n,n} = \sigma_y \otimes I$   \\  
      \hline
    \end{tabular}
    \caption{Mutually anti-commuting set of Bob's observable}
    \label{tab:my_label}
\end{table}

\section{Activation of nonclassicality: setting the stage}
For our purpose, let us assume that Alice and Bob share the state which is a convex mixture of a pure state and color noise as given by
\begin{eqnarray}
\label{pd}
\rho_{2^m} = q |\phi_2 \rangle^{\bigotimes m} \langle \phi_2 |^{\bigotimes m} + (1-q) | 0\rangle \langle 0 | \otimes \frac{\mathbb{I}^{\bigotimes m}_2}{2^{m}}
\end{eqnarray}
Here, $q$ ($0 < q \leq 1$) is the mixing parameter, $m=\lfloor\frac{n}{2}\rfloor$. 
\subsection{Critical value of $q$ for nonlocality and preparation contextuality}
For the set of observables given in Table. I, using mixed entangled state $\rho_{2^m}$, the quantum value of the Bell expression is obtained as 
\begin{align}
\label{maxq1}
\mathcal{B}_{n,Q}= 2^{n-1}\sqrt{n}\ q
\end{align}
By comparing this optimal quantum value for mixed entangled states with local and preparation non-contextual bound, the restriction on mixedness parameter $q$ can be observed. Therefore, the nonlocality and preparation contextuality can be observed for
\begin{eqnarray}
\label{ql}
 q_{nl} > \frac{n\binom{n-1}{\lfloor\frac{n-1}{2}\rfloor} }{ 2^{n-1}\sqrt{n}} ; \ \ 
 q_{pc} > \frac{1}{\sqrt{n}}
\end{eqnarray}
Here, $q_{nl}$ and $ q_{pc}$ are critical values for demonstrating nonlocality and preparation contextuality, respectively.  By comparing $q_{nl}$ and $ q_{pc}$ in Eq.(\ref{ql}), it can be seen that preparation contextuality  is more robust to noise than that of nonlocality, as $q_{nl} \geq q_{pc}$. Therefore, the range of the mixedness parameter $(q_{nl} > q >  q_{pc})$ is the region in which non-classicality in the form of nonlocality cannot be observed, but in the form of preparation contextuality can be revealed. 
In this paper, our aim is to activate the hidden nonlocality and preparation contextuality for the state in Eq.(\ref{pd}) by using the quantum violation of the Bell expression family for different values of $n$. For this we use a local filtering operation on the state and evaluate the quantum value of the Bell expression for a different value of $n$.

\subsection{Local filtering operations}
 Local filtering operation defined as stochastic local operation assisted by classical communication \cite{lli} is used to reveal various hidden quantum correlations.  If before measurement Alice and Bob perform local filters $F_A$ and $F_B$ in their local region then the initial bipartite state $\rho_{d} $ transforms as 
\begin{eqnarray}
\Tilde{\rho}^F_{d} = (F_A \otimes F_B) \rho_{d}  (F_A \otimes F_B)^{\dagger}.
\end{eqnarray}
Here $||F_A||_{\infty} \leq 1$ and $||F_B||_{\infty} \leq 1$, where $||.||_{\infty}$ is Schatten $\infty-$norm. The success probability of transformation of initial state to the filtered state is defined as $N_d = Tr[\Tilde{\rho}^F_{d}]$ and resulting filtered state is obtained as
\begin{eqnarray}
\rho^F_{d} = \frac{\Tilde{\rho}^F_{d}}{N_d}.
\end{eqnarray}
 The concurrence of the filtered state $\rho^F_{d} $ in terms of concurrence of unfiltered state $\rho_{d}$ is defined as
\begin{eqnarray}
C(\rho^F_{d})= \frac{|det(F_A)||det(F_B|}{Tr[(F_A F^{\dagger}_A \otimes F_B F^{\dagger}_B)\rho_{d}]} C(\rho_{d}).
\end{eqnarray}
The filtered state can have a higher concentration of entanglement with respect to initial unfiltered state, as concurrence of filtered state ($C(\rho^F_{d})$) can be increased by considering $|det(F_A)| \neq 0$, $|det(F_B)| \neq 0$ and $|det(F_A)||det(F_AB| > Tr[(F_A F^{\dagger}_A \otimes F_B F^{\dagger}_B)\rho_{d}]$ \cite{jask}. For increasing the concentration of entanglement filtering operation, reduce the noise of the initial shared state (unfiltered state). However, the extent to which noise can be eliminated depends on the type of noise mixed with a pure entangled state \cite{jone}.
We take the mixed entangled state of the form given in Eq. (\ref{pd}) and local filters of Alice and Bob as 
\begin{eqnarray}
\label{df}
F_A = \xi | 0 \rangle \langle 0 | + \sum^{2^{m}-1}_{j=1} | j \rangle \langle j |  \ \ and  \  \ F_B = \delta | 0 \rangle \langle 0 | +  \sum^{2^{m}-1}_{j=1} | j \rangle \langle j | \\ \nonumber
\end{eqnarray}
 respectively \cite{Mat2020}. Here, $0\leq \xi, \delta \leq 1$ with  $\delta =\frac{\xi }{\sqrt{q}}$.






\section{Hidden quantum nonlocality for $n=2,3,4, 5$} 
We demonstrate the hidden nonlocality and preparation contextuality using quantum violation of Bell expressions for the mixed entangled state of the form given in Eq. (\ref{pd}).
\subsection{Activation for the Bell expression for $n=2$ and $n=3$ } 
 Let Alice and Bob perform local filtering operations of the form 
 \begin{eqnarray}
\label{df2}
F_A = \xi | 0 \rangle \langle 0 | + | 1 \rangle \langle 1 | , \ \ \ F_B = \delta | 0 \rangle \langle 0 | +  | 1 \rangle \langle 1 | \\ \nonumber
\end{eqnarray}
on their respective subsystems of the shared state $\rho_2$ in Eq.(\ref{pd}) ($m=\lfloor\frac{n}{2}\rfloor=1$, for $n=2,3$). After the application of local filters $F_A$ and $F_B$, $\rho_2$ transforms as
\begin{eqnarray}
\rho^F_{2} &=&   \frac{1}{N_2} (F_A \otimes F_B) \rho_{2}  (F_A \otimes F_B)^{\dagger} \\ \nonumber   &=& \frac{1}{2 N_2} \bigg[2 q  |\phi_2 \rangle \langle \phi_2 |  +  \sqrt{2}q( \xi^2 \delta^2 -  1) \bigg( ( |\phi_2 )\langle 00 | +| 00 \rangle ( \langle \phi_2 |)\bigg)\\ \nonumber &&  +  (1-q) \xi^2  | 0 1 \rangle \langle 01 |    +  (\xi^2 \delta^2+q(1- 2 \xi \delta))| 00 \rangle \langle 00 |  \bigg],
\end{eqnarray}
where, $N_2 = \frac{1}{2} \left(q+(1-q)\xi ^2+\xi^2 \delta^2 \right)$ is normalization constant. We use this filtered state $\rho^F_{2}$ to derive the quantum value of the Bell expression for $n=2$ and $n=3$.

$\textit{For $n=2$:}$ The quantum value of the Bell expression in Eq.(\ref{nbell1}) for filtered state $\rho^F_{2}$ and mutually anti-commuting set Bob observables listed in Table I is derived as 
\begin{eqnarray}
\label{nbell22}
	\mathcal{B}^F_{2,Q} &=& \frac{2 \sqrt{2} }{N_2} \bigg[ q \xi \delta \bigg]
\end{eqnarray}
The relation between Alice's observables mentioned in Eq.(\ref{abr}) is used to get the final expression. Since, in this case, both local and preparation non-contextual bound of Bell inequality are equal to $2$. The violation of Bell inequality obtained in the range of $1/\sqrt{2} < q \leq 1$ using unfiltered state $\rho_2$ changes to $0.665  < q \leq 1$ at $\xi=0.79$ for filtered state $\rho^F_{2}$. Hence, the range of hidden quantumness obtained after the operation of the local filter is $0.665 < q < 0.707$ \cite{GISIN1996}.\\

$\textit{For $n=3$:}$  In this case quantum value of the Bell expression (Eq.(\ref{nbell1})) for the filtered state $\rho^F_{2}$ and measuring operator given in Table I is obtained as 
\begin{eqnarray}
\label{nbell23}
\mathcal{B}^F_{3,Q} &=&   \frac{4}{N_2\sqrt{3}} \bigg[ 2 q \xi \delta + \frac{q}{2} - \frac{1}{2}(1-q)\xi^2 + \frac{\xi^2 \delta^2}{2}\bigg]
\end{eqnarray}
The detailed calculation is given in Appendix B. Here it should be remarked that, unlike the $n=2$ case, for $n=3$ the local and preparation non-contextual are not same. In this case, nonlocality is observed if $\mathcal{B}^F_{3,Q} > 6$ and quantumness in the form of preparation contextuality is observed if $\mathcal{B}^F_{3,Q} > 4 $. The range of $q$ showing nonlocality and preparation contextuality before the filtering operation are $0.87 < q \leq 1$ and $0.57 < q \leq 1$, respectively. However, after the local filtering operation, the range of the mixedness parameter showing nonlocality changes to $0.86 < q \leq 1$ at $\xi=0.90$ and preparation contextuality changes to  $0.50 < q \leq 1$ at $\xi= 0.70$.
 By comparing the range of mixedness parameters obtained using unfiltered state with filtered state, we can see that both forms of quantumness, quantum nonlocality and preparation contextuality, can be observed for a wider range of mixing parameters. 

\subsection{Activation for the Bell expression for $n=4$ and $n=5$ } 
For $n=4$ and $n=5$, the mixed entangled state shared between Alice and Bob is of the form 
\begin{eqnarray}
\label{pr45}
\rho_{4} = q |\phi_2 \rangle \langle \phi_2 |^{\bigotimes 2} + (1-q)|0 \rangle \langle 0 | \otimes  \frac{\mathbb{I}^{\bigotimes 2}_2}{4},
\end{eqnarray}
 If Alice and Bob locally choose the  filters of the form defined below
\begin{eqnarray}
\label{df0}
F_A &=&  \xi | 0 \rangle \langle 0 | + | 1 \rangle \langle 1 |  + | 2 \rangle \langle 2 |  + | 3 \rangle \langle 3 | , \\ \nonumber  F_B &=&  \delta | 0 \rangle \langle 0 | +  | 1 \rangle \langle 1 | + | 2 \rangle \langle 2 |  + | 3 \rangle \langle 3 |
\end{eqnarray}
 The shared state $\rho_{4} $ after the local operation of $F_A$ and $F_B$ on the respective subsystem of Alice and Bob transform as
\begin{eqnarray}
\rho^F_{4} &=&   \frac{1}{N_4} (F_A \otimes F_B) \rho_{4}  (F_A \otimes F_B)^{\dagger} \\ \nonumber &=&  \frac{1}{4 N_4} \bigg[4 q  |\phi_2 \rangle \langle \phi_2 |^{\bigotimes 2} +  (\xi^2 \delta^2+q(1- 2 \xi \delta))| 00 \rangle \langle 00 |  \\ \nonumber &&  +  (1-q) \xi^2  | 0  \rangle \langle 0  |  \otimes ( | 1 \rangle \langle 1 |  + | 2 \rangle \langle 2 |  + | 3 \rangle \langle 3 | )  \\ \nonumber &&  +  2 q ( \xi^2 \delta^2 -  1) \bigg(  |\phi_2\rangle^{\bigotimes 2} \langle 00 | +| 00 \rangle ( \langle \phi_2 |)^{\bigotimes 2}\bigg) \bigg]
\end{eqnarray}
where, $N_4 =  (\frac{3}{4})[q+ (1-q)\xi^2]  + \frac{ \xi^4}{4 q } $ is the normalization constant.

$\textit{For $n=4$:} $  Using Eq.(\ref{nbell1}), the quantum value of the Bell expression for the filtered state $\rho^F_{4}$ is derived as
\begin{eqnarray}
\label{nbell24}
	\mathcal{B}^F_{4,Q} &=&  \frac{8 q }{N_4}\bigg[ 1 +  \xi \delta  \bigg] 
\end{eqnarray}
In this case nonlocality is observed if $\mathcal{B}_{4,Q} \geq 12 $ and preparation contextuality is observed when $\mathcal{B}_{4,Q} \geq 8$. The range of mixedness parameters for observing nonlocality and preparation contextual obtained using Eq.(\ref{ql}) are $0.75 < q \leq 1$ and $0.5 < q \leq 1$, respectively. However, using $\mathcal{B}^F_{4,Q}$ for the filtered state $\rho^F_{4}$, nonlocality is observed in the range of $0.66<q \leq 1$ and preparation contextuality is observed for any non-zero value of mixedness parameter given $\xi \rightarrow 0$.  \\

$\textit{For $n=5$:} $ Similarly as for $n=4$, the quantum value of Bell expression for $n=5$ is obtained as
\begin{eqnarray}
\label{nbell25}
\mathcal{B}^F_{5,Q} &=&   \frac{16}{N_4\sqrt{5}} \bigg[  2q + 2 q \xi \delta  + \frac{3 q}{4} - \frac{1}{4}(1-q)\xi^2 + \frac{\xi^2 \delta^2}{4}  \bigg]
\end{eqnarray}
 The range of mixedness parameters showing nonlocality and preparation contextual obtained using unfiltered state $\rho_4$ are $0.84 < q \leq 1$ and $0.45 < q \leq 1$ respectively. After local filtering operation nonlocality and preparation contextual are observed in the range of $0.80 < q \leq 1$ at $\xi=0.72$ and $0 < q \leq 1$ for $\xi \rightarrow 0$ respectively.
In the next section, we generalize the application of the local filtering operation for arbitrary $n$
\section{ General Bell expression for filtered state} 
The application of the local filters in Eq. (\ref{df}) on the shared mixed entangled state given in Eq. (\ref{pd}) transform $\rho_{2^m} $  as
\begin{eqnarray}
\rho^F_{2^m} &=& \frac{1}{N_d} \bigg[(F_A \otimes F_B) \rho_{2^m}  (F_A \otimes F_B)^{\dagger}\bigg] \\ \nonumber &=&\frac{1}{N_d 2^{\lfloor\frac{n}{2}\rfloor} } \bigg[q \bigg( \sum^{2^{m}-1}_{j=0}  |j \rangle   |j \rangle \bigg) \bigg( \sum^{2^{m}-1}_{j=0}   \langle j |  \langle  j | \bigg)\\ \nonumber &+& \sqrt{q}( \xi^2-  \sqrt{q}) \bigg[ \bigg( \sum^{2^{m}-1}_{j=0}  |j \rangle   |j \rangle \bigg) \langle 00 |+ | 00 \rangle \bigg( \sum^{2^{m}-1}_{j=0}   \langle j |  \langle  j |  \bigg) \bigg]  \\ \nonumber &+&    (1-q) \xi^2  | 0 \rangle \langle 0 | \otimes \sum^{2^{m}-1}_{j=1}  |j \rangle \langle j | + \bigg( \frac{(1-q)\xi^4 }{q}  \\ \nonumber &+&  ( \xi^2-\sqrt{q} )^2\bigg) | 00 \rangle \langle 00 | \bigg]
\end{eqnarray}
where, $N_d = [q+ (1-q)\xi^2] (1-\frac{1}{2^{\lfloor\frac{n}{2}\rfloor}}) + \frac{ \xi^4}{q 2^{\lfloor\frac{n}{2}\rfloor}} $ is the normalization constant.
\begin{figure}[htp]
  \includegraphics[width=4.2cm]{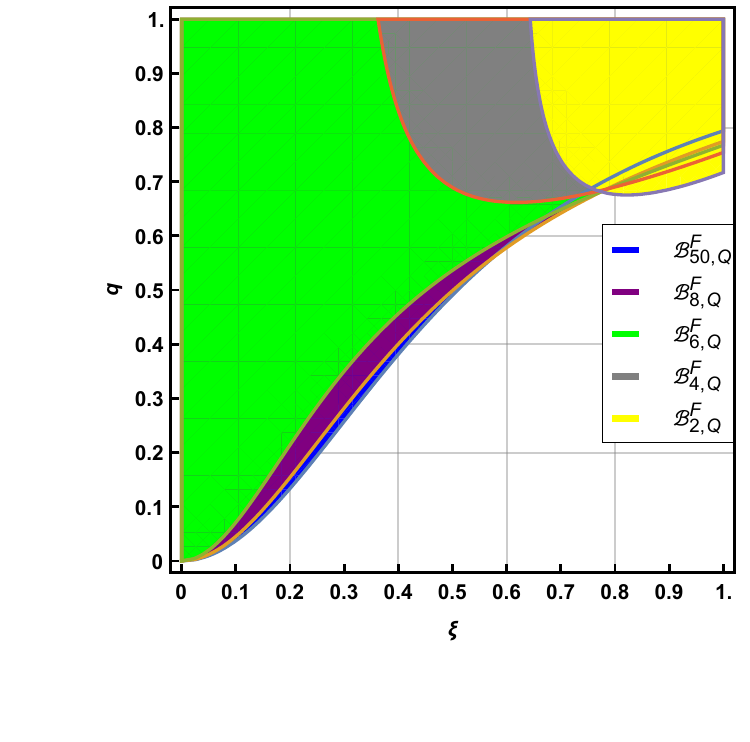}
   \includegraphics[width=4.2cm]{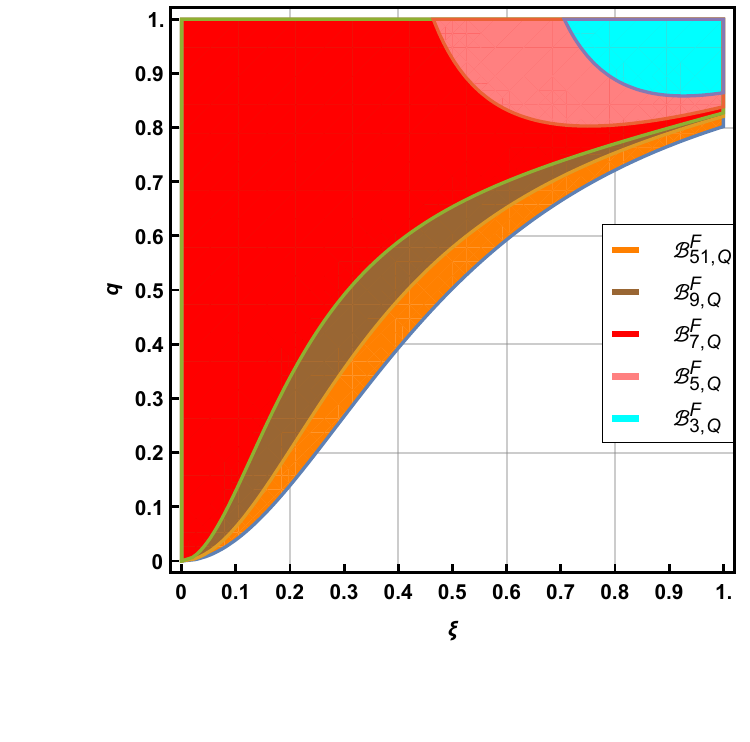}
\caption{Shaded region show the nonlocal region formed when $\mathcal{B}^F_{n,Q} > n\binom{n-1}{\lfloor\frac{n-1}{2}\rfloor}$ (local bound) after local filtering operation.}
    \label{fig:BLP}
\end{figure}

\begin{figure}[htp]
  \includegraphics[width=4.2cm]{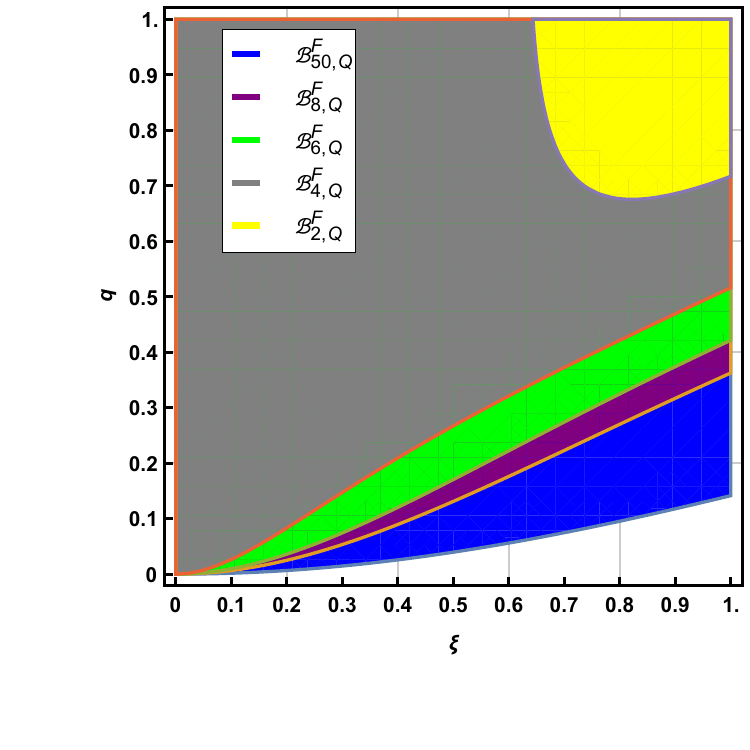}
   \includegraphics[width=4.2cm]{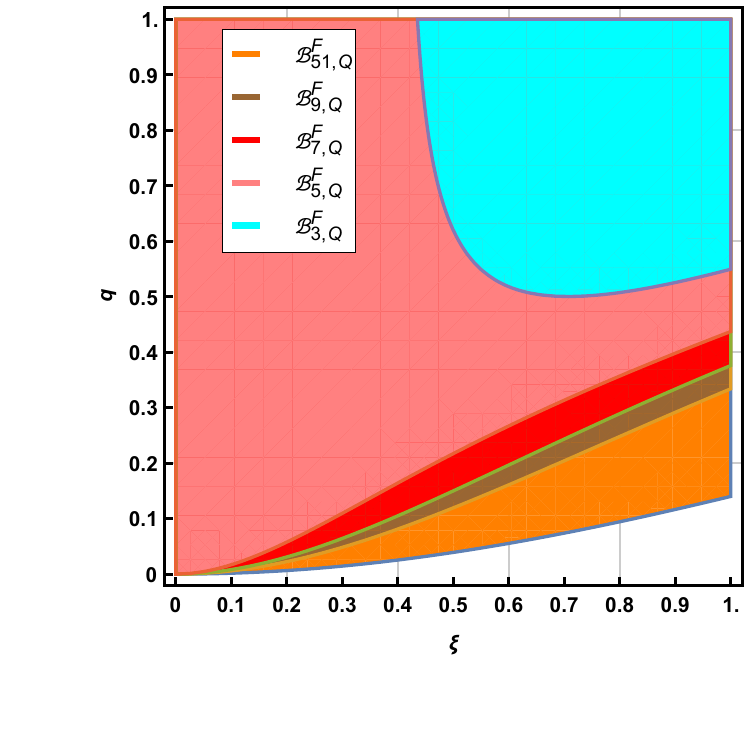}
\caption{Shaded region shows preparation contextual region formed when $\mathcal{B}^F_{n,Q} > 2^{n-1}$ (PNC bound) after local filtering operation.}
    \label{fig:BLP}
\end{figure}
\subsection{ Even $n$ ($n=2v$), $v=1,2,3,.....,v$} 
The quantum value of the Bell expression in Eq.(\ref{nbell1}), using a set of observables in Table.1 and filtered state for even $n$ is derived as
\begin{eqnarray}
\label{nbelle}
	\mathcal{B}^F_{2v,Q} = \frac{2^{2v-1} \sqrt{2v}}{N_d} \bigg[ \frac{q \xi \delta}{2^{v-1}}+\sum_{v=2}^{\lfloor\frac{n}{2}\rfloor} \frac{q}{2^{v-1}} \bigg]
\end{eqnarray}
Detailed calculation of $\mathcal{B}^F_{2v}$ is given in Appendix B. 

\subsection{ Odd $n$ ($n=2v+1$), $v=1,2,3,.....,v$} Similarly for odd $n$, using the set of Bob's observables given in Table 1, the quantum value of the Bell expression is obtained as
\begin{eqnarray}
\label{nbello}
	\mathcal{B}^F_{2v+1,Q} &=& \frac{2v 2^{2v}}{N_d \sqrt{2v+1}}\bigg[ \frac{q \xi \delta}{2^{v-1}}+\sum_{v=2}^{\lfloor\frac{n}{2}\rfloor} \frac{q}{2^{v-1}} \bigg] \\ \nonumber &+& \frac{2^{2v}}{\sqrt{2v+1}N_d}\bigg[ \frac{ \xi^2 \delta^2-(1-q)\xi^2}{2^{v}}+\sum_{v=1}^{\lfloor\frac{n}{2}\rfloor} \frac{q}{2^{v}} \bigg]
\end{eqnarray}
A detailed derivation of  $\mathcal{B}^F_{2v+1,Q}$ is given in Appendix B. The range of $q$ that reveals nonlocality and preparation contextuality is obtained if $\mathcal{B}^F_{n,Q} >  n\binom{n-1}{\lfloor\frac{n-1}{2}\rfloor}$ and $\mathcal{B}^F_{n,Q} > 2^{n-1}$  respectively. In Fig.1 we have plotted the nonlocal region separately for even and odd $n$ against the mixing parameter $q$ and parameter $\xi$ used in local filtering operation. From Fig. 1 we can see that nonlocality is observed for any non-zero value of $q$ if $n\geq 6$ for even $n$. However, in case of odd $n$ nonlocality is observed for the whole range mixedness parameter if $n\geq 7$. The region plot showing preparation contextuality for even and odd $n$ is plotted in Fig. 2. It can be observed from Fig. 2 that preparation contextuality can be observed for any non-zero value of $q$ for even and odd $n$ if $n\geq 4$ and $n\geq 5$ respectively.

\begin{figure}[htp]
   \includegraphics[width=8cm]{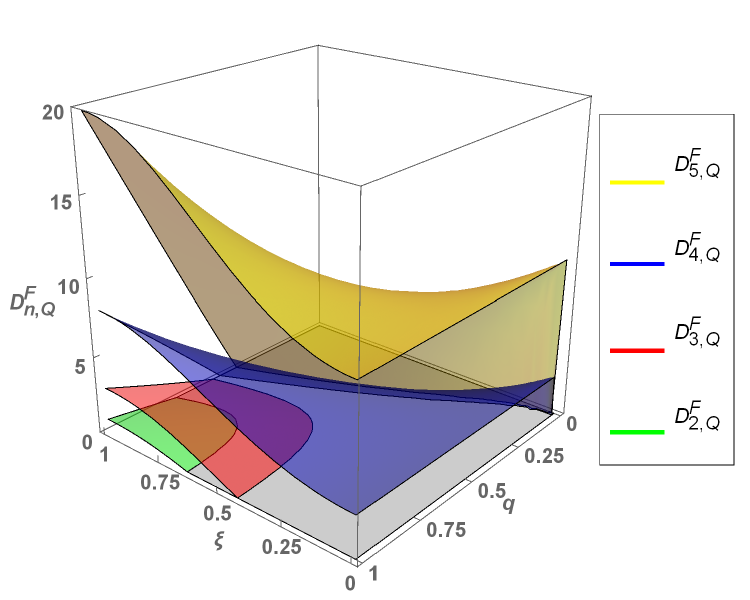}
\caption{Difference of quantum value of Bell expression obtained using filtered state and preparation non-contextual bound ($D^F_{n,Q}=\mathcal{B}^F_{n,Q} - 2^{n-1}$) for $n=2,3,4,5$ are plotted with respect to $q$ and $\xi$.}
    \label{fig:BLP}
\end{figure}

In order to see the departure of quantum value from preparation non-contextual bound after the operation of local filters in Fig.3, we have plotted the difference of quantum value of Bell expression obtained using filtered state and preparation non-contextual bound ($D^F_{n,Q}=\mathcal{B}^F_{n,Q} - 2^{n-1}$) for different values of $n$. We found that for $n \geq 4$, the difference, $D^F_{n,Q}$ is positive for any nonzero value of $q$. This gives direct indication of obtaining quantum advantanges in success probability of $n$-bit PORAC communication game for any non-zero value of mixing parameter if $n \geq 4$.

 \section{Summary and discussion}
In summary, we have demonstrated both the activation of hidden nonlocality and preparation contextuality using the quantum violation of a family of Bell inequalities. For our demonstration, we used a specific class of local filtering operation as a tool. We have shown the activation of both forms of quantumness by considering a mixture of pure states and color noise. We found that non-locality can be activated for any non-zero value of the mixedness parameter if $n \geq 6$. Next, we found that preparation contextuality can be activated for the whole range of mixedness parameters for $n \geq 4$. 

The activation of hidden nonlocality and preparation contextuality we have demonstrated is applicable for two-party Alice and Bob performing $2^{n-1}$ and $n$ dichotomic measurements on their respective subsystem. Our next goal along this line is to study the hidden nonlocality and preparation contextuality for multi-outcome and multiparty scenarios.

\section*{Aknowledgement} 
Authors are indebted to Dr. Alok Kumar Pan (IIT Hyderabad, India) for fruitful discussions. A.K acknowledges the Research Associateship from Indian Institute of Science Education and Research, Berhampur, India.

\begin{widetext}
\appendix
\section{Effect of local filters on mixed entangled state for $n$-bit case.}
If Alice and Bob apply the local filters defined in Eq. (\ref{df}) on the respective subsystems of their shared mixed entangled state $\rho_{2^m} $ given in Eq. (\ref{pd}), then unfiltered state $\rho_{2^m} $ transforms as 
\begin{eqnarray}
\label{a}
\rho^F_{2^m} &=& \frac{1}{N_d} \bigg[(F_A \otimes F_B) \rho_{2^m}  (F_A \otimes F_B)^{\dagger}\bigg] \\ \nonumber &=& \frac{1}{N_d} \bigg[q (F_A \otimes F_B) |\phi_{2^m} \rangle \langle \phi_{2^m} |  (F_A \otimes F_B)^{\dagger} + \frac{1-q}{d} (F_A \otimes F_B) | 0 \rangle \langle 0 | \otimes \mathbb{I}  (F_A \otimes F_B)^{\dagger} \bigg] \\ \nonumber &=&   \frac{q}{N_d 2^{\lfloor\frac{n}{2}\rfloor}}\bigg[ \delta \xi | 00 \rangle +  \sum^{2^{\lfloor\frac{n}{2}\rfloor}-1}_{j=1}  |j \rangle_A   |j \rangle_B \bigg]\bigg[ \delta \xi \langle 00 | +  \sum^{2^{\lfloor\frac{n}{2}\rfloor}-1}_{j=1}   \langle j |  \langle  j | \bigg] +  \bigg( \frac{1-q}{2^{\lfloor\frac{n}{2}\rfloor}} \bigg)  \xi^2  | 0 \rangle \langle 0 | \otimes \bigg( \delta^2 | 0 \rangle \langle 0 | +  \sum^{2^{\lfloor\frac{n}{2}\rfloor}-1}_{j=1}  |j \rangle \langle  j|  \bigg)  \\ \nonumber  &=&   \frac{q}{N_d 2^{\lfloor\frac{n}{2}\rfloor}}\bigg[ (\delta \xi -1 )| 00 \rangle +  \sum^{2^{\lfloor\frac{n}{2}\rfloor}-1}_{j=0}  |j \rangle   |j \rangle \bigg]\bigg[ (\delta \xi-1 ) \langle 00 | +  \sum^{2^{\lfloor\frac{n}{2}\rfloor}-1}_{j=0}   \langle j |  \langle  j | \bigg] + \bigg( \frac{1-q}{2^{\lfloor\frac{n}{2}\rfloor}} \bigg)  \xi^2  | 0 \rangle \langle 0 | \otimes \bigg( \delta^2 | 0 \rangle \langle 0 | +  \sum^{2^{\lfloor\frac{n}{2}\rfloor}-1}_{j=0}  |j \rangle \langle  j |  \bigg) \\ \nonumber  &=& \frac{1}{N_d 2^{\lfloor\frac{n}{2}\rfloor}} \bigg[ q(\delta \xi-1 )^2 | 00 \rangle \langle 00 |+ q(\delta \xi-1 ) \bigg( \sum^{2^{\lfloor\frac{n}{2}\rfloor}-1}_{j=0}  |j \rangle   |j \rangle  \langle 00 | +  | 00 \rangle   \sum^{2^{\lfloor\frac{n}{2}\rfloor}-1}_{j=0}   \langle j |  \langle  j | \bigg)   + q \bigg( \sum^{2^{\lfloor\frac{n}{2}\rfloor}-1}_{j=0}  |j \rangle   |j \rangle \bigg) \bigg( \sum^{2^{\lfloor\frac{n}{2}\rfloor}-1}_{j=0}   \langle j |  \langle  j | \bigg) \\ \nonumber &&  +  (1-q) \delta^2 \xi^2 | 0 \rangle \langle 0 | \otimes  | 0 \rangle \langle 1 | + (1-q) \xi^2  | 0 \rangle \langle 0 | \otimes \sum^{2^{\lfloor\frac{n}{2}\rfloor}-1}_{j=1}  |j \rangle \langle j |  \bigg] \\ \nonumber &=& \frac{1}{N_d 2^{\lfloor\frac{n}{2}\rfloor} } \bigg[q \bigg( \sum^{2^{\lfloor\frac{n}{2}\rfloor}-1}_{j=0}  |j \rangle   |j \rangle \bigg) \bigg( \sum^{2^{\lfloor\frac{n}{2}\rfloor}-1}_{j=0}   \langle j |  \langle  j | \bigg)+ \sqrt{q}( \xi^2-  \sqrt{q}) \bigg[ \bigg( \sum^{2^{\lfloor\frac{n}{2}\rfloor}-1}_{j=0}  |j \rangle   |j \rangle \bigg) \langle 00 | +| 00 \rangle \bigg( \sum^{2^{\lfloor\frac{n}{2}\rfloor}-1}_{j=0}   \langle j |  \langle  j |  \bigg) \bigg]  \\ \nonumber &&   + (1-q) \xi^2  | 0 \rangle \langle 0 | \otimes \sum^{2^{\lfloor\frac{n}{2}\rfloor}-1}_{j=1}  |j \rangle \langle j | + \bigg( \frac{(1-q)\xi^4 }{q}  + ( \xi^2-\sqrt{q} )^2\bigg) | 00 \rangle \langle 00 | \bigg]
\end{eqnarray}
where, $N_d =Tr\bigg[(F_A \otimes F_B) \rho_{d}  (F_A \otimes F_B)^{\dagger}\bigg] = [q+ (1-q)\xi^2] (1-\frac{1}{2^{\lfloor\frac{n}{2}\rfloor}}) + \frac{ \xi^4}{q 2^{\lfloor\frac{n}{2}\rfloor}} $ is the normalization constant and last line is obtained using $\delta =\frac{\xi }{\sqrt{q}}$.

\section{Quantum value of general Bell expression for filtered state}
\subsection{Detailed calculation of Bell expression of even $n$ }
For $n=2$, using relation between Alice and Bob observables given in Eq.(\ref{abr})
Bell expression in Eq.(\ref{nbell1}) reduces to
\begin{eqnarray}
	\mathcal{B}^F_{2,Q} &=& \sqrt{2} \bigg[ \langle B_{2,1}\otimes B_{2,1} \rangle + \langle B_{2,2}\otimes B_{2,2} \rangle \bigg] 
\end{eqnarray}
Using mutually anti-commuting set of operators of Bob in Table. I for the filtered state $\rho^F_{2}$ we get
\begin{align}
 &&\langle B_{2,1}\otimes B_{2,1} \rangle=   \langle B_{2,2}\otimes B_{2,2} \rangle  = \frac{\delta  \xi  q}{N_2}
\end{align}
 Then the quantum value of the Bell expression is obtained as
\begin{eqnarray}
	\mathcal{B}^F_{2,Q} &=& \frac{2 \sqrt{2} }{N_2} \bigg[ q \xi \delta \bigg]
\end{eqnarray}

For $n=4$ using relation between Alice and Bob observables Bell expression can be rewritten as
\begin{eqnarray}
	\mathcal{B}^F_{4,Q} &=& 4 \bigg[ \langle B_{4,1}\otimes B_{4,1} \rangle + \langle B_{4,2}\otimes B_{4,2} \rangle + \langle B_{4,3}\otimes B_{4,3} \rangle + \langle B_{4,4}\otimes B_{4,4} \rangle\bigg]
\end{eqnarray}
In this case joint expectation value involved in the Bell expression for filtered state $\rho^F_{4}$ is obtained as
\begin{align}
\label{e4}
\langle B_{4,1}\otimes B_{4,1} \rangle = \langle B_{4,2}\otimes B_{4,2} \rangle  = \langle B_{4,3}\otimes B_{4,3} \rangle  = \langle B_{4,3}\otimes B_{4,3} \rangle = \frac{1}{N_4}  \bigg[\frac{\delta  \xi  q}{2}+\frac{q}{2} \bigg] 
\end{align}
Using Eq.({\ref{e4}}), quantum value of the Bell expression is obtained as
\begin{eqnarray}
	\mathcal{B}^F_{4,Q} &=& \frac{16}{N_4} \bigg[ \frac{\delta  \xi  q}{2}+\frac{q}{2}  \bigg] 
\end{eqnarray}
Using same relation between Alice and Bob observables for $n=6$, the form of Bell expression is
\begin{eqnarray}
	\mathcal{B}^F_{6,Q} &=&  \frac{32}{\sqrt{5}}  \bigg[ \langle B_{6,1}\otimes B_{6,1} \rangle + \langle B_{6,2}\otimes B_{6,2} \rangle + \langle B_{6,3}\otimes B_{6,3} \rangle  + \langle B_{6,4}\otimes B_{6,4} \rangle + \langle B_{6,5}\otimes B_{6,5}  \rangle  + \langle B_{6,6}\otimes B_{6,6}  \rangle \bigg] 
\end{eqnarray}
If shared state is  $\rho^F_{8}$ we have
\begin{align}
  \langle B_{6,1}\otimes B_{6,1} \rangle = \langle B_{6,2}\otimes B_{6,2} \rangle = \langle B_{6,3}\otimes B_{6,3} \rangle =\langle B_{6,4}\otimes B_{6,4} \rangle = \langle B_{6,5}\otimes B_{6,5} \rangle = \langle B_{6,6}\otimes B_{6,6} \rangle =  \frac{1}{N_8} \bigg[ \frac{\delta  \xi  q}{4}+\frac{3 q}{4} \bigg]
 \end{align}
and quantum value of the Bell expression is given by
\begin{eqnarray}
\mathcal{B}^F_{6,Q} &=&  \frac{192}{N_8\sqrt{5}} \bigg[ \frac{\delta  \xi  q}{4}+\frac{3 q}{4} \bigg]
\end{eqnarray}
Similarly, using the induction method, expectation value of joint correlations of Alice and Bob observable in Bell expression for even $n$ ($n=2v$) is obtained as 
\begin{align}
\langle B_{2v,y}\otimes B_{2v,y} \rangle = \frac{1}{N_d} \bigg[ \frac{q \xi \delta}{2^{v-1}}+\sum_{v=2}^{\lfloor\frac{n}{2}\rfloor} \frac{q}{2^{v-1}} \bigg]
\end{align}
The corresponding Bell expression is derived as
\begin{eqnarray}
\label{ev2n}
    \mathcal{B}^F_{2v,Q} = \frac{2^{2v-1}}{ \sqrt{2v}} \sum_{y=1}^{2v} \langle B_{2v,y}\otimes B_{2v,y} \rangle   = \frac{2^{2v-1} \sqrt{2v}}{N_d} \bigg[ \frac{q \xi \delta}{2^{v-1}}+\sum_{v=2}^{\lfloor\frac{n}{2}\rfloor} \frac{q}{2^{v-1}} \bigg]
\end{eqnarray}

\subsection{Detailed calculation of Bell expression of odd $n$ }
For $n=3$, following Eq.(\ref{abr}) Bell expression in Eq.(\ref{nbell1}) reduces to
\begin{eqnarray}
	\mathcal{B}^F_{3,Q} &=&  \frac{4}{\sqrt{3}} \bigg[ \langle B_{3,1}\otimes B_{3,1} \rangle + \langle B_{3,2}\otimes B_{3,2} \rangle + \langle B_{3,3}\otimes B_{3,3} \rangle\bigg] 
\end{eqnarray}
By comparing the set anti-commuting set of Bob observables in Table.1 for $n=2$ and $n=3$ we can see that first two terms of the Bell expression, $\langle B_{3,1}\otimes B_{3,1} \rangle = \langle B_{3,2}\otimes B_{3,2} \rangle= \frac{\delta  \xi q}{N_2}$ as obtained for $\langle B_{2,y=1,2}\otimes B_{2,y=1,2} \rangle$. However, the last term is obtained as
\begin{align}
   \langle B_{3,3}\otimes B_{3,3} \rangle=\langle\sigma_y \otimes \sigma_y \rangle = \frac{1}{N_2}\bigg[\frac{\delta ^2 \xi ^2}{2}+\frac{1}{2} \xi ^2 (q-1)+\frac{q}{2} \bigg]
\end{align}
Then the quantum value of the Bell expression using the filtered state $\rho^F_{2}$ for $n=3$ is obtained as
\begin{eqnarray}
	\mathcal{B}^F_{3,Q} &=&    \frac{4}{N_2\sqrt{3}} \bigg[   2 q \xi \delta + \frac{q}{2} - \frac{1}{2}(1-q)\xi^2 + \frac{\delta^2\xi^2}{2}\bigg]
    \end{eqnarray}
    In case of $n=5$, the form of Bell expression is\\
\begin{eqnarray}
	\mathcal{B}^F_{5,Q} &=&  \frac{16}{\sqrt{5}} \bigg[ \langle B_{5,1}\otimes B_{5,1} \rangle + \langle B_{5,2}\otimes B_{5,2} \rangle + \langle B_{5,3}\otimes B_{5,3} \rangle + \langle B_{5,4}\otimes B_{5,4} \rangle + \langle B_{5,5}\otimes B_{5,5} \rangle \bigg]
\end{eqnarray}
For the filtered state $\rho^F_{2}$ the quantum value of first four term in this case are equal and same as obtained for $n=4$
\begin{align}
  \langle B_{5,y=1,2,3,4}\otimes B_{5,y=1,2,3,4} \rangle = \frac{1}{N_4} \bigg[\frac{\delta  \xi  q}{2}+\frac{q}{2}\bigg]  
\end{align}
 The quantum value of $\langle B_{5,5}\otimes B_{5,5} \rangle$ is obtained as
\[ \langle B_{5,5}\otimes B_{5,5} \rangle  = \frac{1}{N_2} \bigg[  \frac{\delta ^2 \xi ^2}{4}+\frac{1}{4} \xi ^2 (q-1)+\frac{3 q}{4}\bigg] \]\\
The quantum value of the Bell expression for $n=5$ is obtained as
\begin{eqnarray}
	\mathcal{B}^F_{5,Q} &=&  \frac{64}{N_2 \sqrt{5}} \bigg[ \frac{\delta  \xi  q}{2}+\frac{q}{2}  \bigg] + \frac{16}{N_2 \sqrt{5}} \bigg[\frac{1}{2}\bigg(\frac{\delta ^2 \xi ^2}{2}+\frac{1}{2} \xi ^2 (q-1)\bigg) + \frac{3q}{4}  \bigg] 
\end{eqnarray}
Similarly for  $n=2v+1$, using the induction method for $\langle B_{2v,y}\otimes B_{2v,y} \rangle$  and quantum value of Bell expression for $n=2v$ given in Eq.(\ref{ev2n}) we get
\begin{eqnarray}
    \mathcal{B}^F_{2v+1,Q} = \frac{2v 2^{2v}}{N_d \sqrt{2v+1}}\bigg[ \frac{q \xi \delta}{2^{v-1}}+\sum_{v=2}^{\lfloor\frac{n}{2}\rfloor} \frac{q}{2^{v-1}} \bigg]   + \frac{2^{2v}}{\sqrt{2v+1}N_d}\bigg[ \frac{ \xi^2 \delta^2-(1-q)\xi^2}{2^{v}}+\sum_{v=1}^{\lfloor\frac{n}{2}\rfloor} \frac{q}{2^{v}} \bigg]
\end{eqnarray}

\end{widetext}

	\end{document}